\documentstyle [12pt] {article}

\catcode`@=12
\begin{document}
\thispagestyle{empty}
\begin{flushright} UCRHEP-T107\\TRI-PP-93-17\\May 1993\\hep-ph/9305230\
\end{flushright}
\vspace{0.5in}
\begin{center}
{\Large \bf New Supersymmetric Option\\for Two Higgs Doublets\\}
\vspace{1.0in}
{\bf Ernest Ma\\}
{\sl Department of Physics\\}
{\sl University of California\\}
{\sl Riverside, California 92521\\}
\vspace{0.3in}
{\bf Daniel Ng\\}
{\sl TRIUMF, 4004 Wesbrook Mall\\}
{\sl Vancouver, British Columbia\\}
{\sl Canada V6T 2A3\\}
\vspace{1.0in}
\end{center}
\begin{abstract}\
If the standard electroweak gauge model is embedded in a larger theory which
is supersymmetric and the latter breaks down to the former at some mass
scale, then the reduced Higgs potential at the electroweak mass scale may
differ from that of the well-known minimal supersymmetric extension.
Specifically, if the larger theory is based on $\rm {SU(2)_L \times SU(2)_R
\times U(1)}$, an interesting alternative exists for two Higgs doublets.
\end{abstract}

\newpage
\baselineskip 24pt

The most studied extension of the standard SU(2) $\times$ U(1) electroweak
gauge model is that of supersymmetry with the smallest necessary particle
content.\cite{hk}
In this minimal supersymmetric standard model (MSSM), there are two scalar
doublets $\Phi_1 = (\phi_1^+, \phi_1^0)$ and $\Phi_2 = (\phi_2^+, \phi_2^0)$,
with Yukawa interactions $\overline {(u,d)}_L d_R \Phi_1$ and $\overline
{(u,d)}_L u_R \tilde \Phi_2$ respectively, where $\tilde \Phi_2 = i \sigma_2
\Phi_2^* = (\overline {\phi_2^0}, - \phi_2^-)$.  The Higgs potential
\begin{eqnarray}
V &=& \mu_1^2 \Phi_1^\dagger \Phi_1 + \mu_2^2 \Phi_2^\dagger \Phi_2 +
\mu_{12}^2 (\Phi_1^\dagger \Phi_2 + \Phi_2^\dagger \Phi_1) \nonumber \\
&+& {1 \over 2} \lambda_1 (\Phi_1^\dagger \Phi_1)^2 + {1 \over 2} \lambda_2
(\Phi_2^\dagger \Phi_2)^2 + \lambda_3 (\Phi_1^\dagger \Phi_1) (\Phi_2^\dagger
\Phi_2) \nonumber \\ &+& \lambda_4 (\Phi_1^\dagger \Phi_2) (\Phi_2^\dagger
\Phi_1) + {1 \over 2} \lambda_5 (\Phi_1^\dagger \Phi_2)^2 + {1 \over 2}
\lambda_5^* (\Phi_2^\dagger \Phi_1)^2,
\end{eqnarray}
is subject to the constraints
\begin{equation}
\lambda_1 = \lambda_2 = {1 \over 4} (g_1^2 + g_2^2),~~\lambda_3 = -{1 \over 4}
g_1^2 + {1 \over 4} g_2^2,~~\lambda_4 = -{1 \over 2} g_2^2,~~\lambda_5 = 0,
\end{equation}
where $g_1$ and $g_2$ are the U(1) and SU(2) gauge couplings of the standard
model respectively.  Hence there are only two unknown parameters in this
sector and they are usually taken to be $\tan\beta \equiv v_2/v_1$, the
ratio of the two scalar vacuum expectation values, and $m_A$, the mass of
its one physical pseudoscalar particle.  Numerous phenomenological
studies\cite{bcpetc} have been made in its name.

It is generally believed that given the gauge group SU(2) $\times$ U(1) and
the requirement of supersymmetry, the quartic scalar couplings of Eq. (1)
must necessarily be given by Eq. (2).  This is actually not the case because
the SU(2) $\times$ U(1) gauge symmetry may be a remnant\cite{mn} of a
larger symmetry which is broken at a higher mass scale {\underline {together
with the supersymmetry}}.  The structure of the Higgs potential is then
determined by the scalar particle content needed to precipitate the proper
spontaneous symmetry breaking and to render massive the assumed fermionic
content of the larger theory.  Furthermore, the quartic scalar couplings are
related to the gauge couplings of the larger theory {\underline {as well as
other couplings}} appearing in its superpotential.  At the electroweak energy
scale, the reduced Higgs potential may contain only two scalar doublets, but
their quartic couplings may not be those of the MSSM.  In particular, we
consider in the following a left-right supersymmetric model based on E$_6$
particle content proposed some years ago\cite{mbhm} and show that its
reduced Higgs potential $V$ for two scalar doublets is given by
\begin{eqnarray}
\lambda_1 &=& {1 \over 4} \left( 1 + {{4 f^2} \over g_2^2} \right) \left[
g_1^2 + g_2^2 - 4 f^2 \left( 1 - {g_1^2 \over g_2^2} \right) \right], \\
\lambda_2 &=& {1 \over 2} g_2^2 + {1 \over 4} (g_1^2 - g_2^2) \left( 1 -
{{4 f^2} \over g_2^2} \right)^2, \\ \lambda_3 &=& {1 \over 4} g_2^2 -
{1 \over 4} \left( 1 - {{4 f^2} \over g_2^2} \right) \left[ g_1^2 - 4 f^2
\left( 1 - {g_1^2 \over g_2^2} \right) \right], \\ \lambda_4 &=& f^2 -
{1 \over 2} g_2^2,~~~\lambda_5 = 0,
\end{eqnarray}
where $f$ is a coupling in the superpotential of the larger theory and has
no analog in the MSSM.  In the limit $f = 0$, it is easily seen from the
above that the MSSM conditions, {\it i.e.} Eq. (2), are recovered as expected.
The general requirement that $V$ be bounded from below puts an upper bound on
$f^2$, namely
\begin{equation}
f^2 \le {1 \over 4} (g_1^2 + g_2^2) \left( 1 - {g_1^2 \over g_2^2}
\right)^{-1}.
\end{equation}
The saturation of this upper bound turns out to imply that the left-right
symmetry of the larger theory is not broken by the soft terms of the Higgs
potential which break the supersymmetry.\cite{mbhm}  In that case,
\begin{eqnarray}
\lambda_1 &=& 0,~~~\lambda_2 ~=~ {1 \over 2} g_2^2 - {g_1^4 \over {g_2^2 -
g_1^2}}, \nonumber \\ \lambda_3 &=& {1 \over 4} g_2^2 - {{g_1^2 g_2^2} \over
{2(g_2^2 - g_1^2)}} ~=~ - \lambda_4,~~~ \lambda_5 = 0.
\end{eqnarray}
The lesson we learn here is that even if supersymmetry exists and there are
only two scalar doublets at the electroweak energy scale, the corresponding
Higgs potential is not necessarily that of the MSSM.

We now describe our model.  The gauge group is $\rm SU(2)_L \times SU(2)_R
\times U(1)$ but with an unconventional assignment of fermions.\cite{mbhm}
An exotic quark $h$ of electric charge $-1/3$ is added so that $(u,d)_L$
transforms as (2,1,1/6), $(u,h)_R$ as (1,2,1/6), whereas both $d_R$ and $h_L$
are singlets $(1,1,-1/3)$.  There are two scalar doublets $\Phi_{1,2}$ and
a bidoublet
\begin{equation}
\eta = \left( \begin{array} {c@{\quad}c} \overline {\eta_1^0} & \eta_2^+ \\
-\eta_1^- & \eta_2^0 \end{array} \right)
\end{equation}
transforming as (2,1,1/2), (1,2,1/2), and (2,2,0) respectively.  The Yukawa
interactions are such\cite{mbhm} that $m_h$ comes from $\langle \phi_2^0
\rangle = v_2$, $m_d$ comes from $\langle \phi_1^0 \rangle = v_1$, and $m_u$
comes from $\langle \eta_1^0 \rangle = u_1$.  The part of the Higgs potential
related to the gauge interactions through supersymmetry is given by\cite{mbhm}
\begin{eqnarray}
V_D &=& {1 \over 8} G_1^2 (\Phi_1^\dagger \Phi_1 - \Phi_2^\dagger \Phi_2)^2
\nonumber \\ &+& {1 \over 8} G_2^2 [ (\Phi_1^\dagger \Phi_1)^2 +
(\Phi_2^\dagger \Phi_2)^2 + 2 (Tr~ \eta^\dagger \eta)^2 - 2 (Tr~ \eta^\dagger
\tilde \eta) (Tr~ {\tilde \eta}^\dagger \eta) \nonumber \\ &~& -
2 (\Phi_1^\dagger \Phi_1 + \Phi_2^\dagger \Phi_2) (Tr~ \eta^\dagger \eta) +
4 (\Phi_1^\dagger \eta \eta^\dagger \Phi_1 + \Phi_2^\dagger \eta^\dagger
\eta \Phi_2) ],
\end{eqnarray}
where $G_1$ is the U(1) gauge coupling and $G_2$ is the coupling of both
$\rm SU(2)_L$ and $\rm SU(2)_R$, with
\begin{equation}
\tilde \eta \equiv \sigma_2 \eta^* \sigma_2 = \left( \begin{array}
{c@{\quad}c} \overline {\eta_2^0} & \eta_1^+ \\ -\eta_2^- & \eta_1^0
\end{array} \right).
\end{equation}
Now the superpotential of this model also contains a cubic term linking
the three superfields corresponding to $\Phi_{1,2}$ and $\eta$ as already
discussed by Babu {\it et al.}\cite{mbhm}  Its contribution to the Higgs
potential is given by
\begin{equation}
V_F = f^2 [(\Phi_1^\dagger \Phi_1)(\Phi_2^\dagger \Phi_2) + (\Phi_1^\dagger
\Phi_1 + \Phi_2^\dagger \Phi_2) (Tr~ \eta^\dagger \eta) - \Phi_1^\dagger \eta
\eta^\dagger \Phi_1 - \Phi_2^\dagger \eta^\dagger \eta \Phi_2].
\end{equation}
To break the gauge symmetry spontaneously, we add soft terms which also break
the supersymmetry:
\begin{equation}
V_{soft} = m_1^2 \Phi_1^\dagger \Phi_1 + m_2^2 \Phi_2^\dagger \Phi_2
+ m_3^2 (Tr~ \eta^\dagger \eta) - fA (\Phi_1^\dagger \tilde \eta \Phi_2 +
\Phi_2^\dagger {\tilde \eta}^\dagger \Phi_1).
\end{equation}
The sum is then
\begin{eqnarray}
V &=& V_{soft} + {1 \over 8} (G_1^2 + G_2^2) [ (\Phi_1^\dagger \Phi_1)^2 +
(\Phi_2^\dagger \Phi_2)^2] \nonumber \\ &+& {1 \over 4} G_2^2 [ (Tr~
\eta^\dagger \eta)^2 - (Tr~ \eta^\dagger \tilde \eta) (Tr~ {\tilde
\eta}^\dagger
\eta) ] + \left( f^2 - {1 \over 4} G_2^2 \right)
(\Phi_1^\dagger \Phi_1 + \Phi_2^\dagger \Phi_2) (Tr~ \eta^\dagger \eta)
\nonumber \\ &-& \left( f^2 - {1 \over 2} G_2^2 \right) (\Phi_1^\dagger \eta
\eta^\dagger \Phi_1 + \Phi_2^\dagger \eta^\dagger \eta \Phi_2)
+ \left( f^2 - {1 \over 4} G_1^2 \right) (\Phi_1^\dagger \Phi_1)
(\Phi_2^\dagger \Phi_2).
\end{eqnarray}
Note that $V$ is invariant also under a global U(1) transformation related
to lepton number as a consequence of the theory's E$_6$ superstring
antecedent and it remains unbroken after spontaneous breaking of the
gauge symmetry.\cite{mbhm}

Consider now the breaking of $\rm SU(2)_L \times SU(2)_R \times U(1)$ down to
the standard $\rm SU(2)_L \times U(1)_Y$.  This is accomplished with
$\langle \phi_2^0 \rangle = v_2 \neq 0$.  Three of the four degrees of
freedom contained in $\Phi_2$ are then absorbed into the three massive
vector gauge bosons W$_R^{\pm}$ and Z', and the remaining neutral physical
Higgs boson ($\sqrt 2$Re$\phi_2^0$) picks up a mass equal to the square root
of $(G_1^2 + G_2^2)v_2^2/2$.  Concurrently, the exotic $h$ quarks and
the $\eta_2$ components of the scalar bidoublet become heavy at the same
mass scale.  The reduced Higgs potential involving only the $(\phi_1^+,
\phi_1^0)$ and $(\eta_1^+, \eta_1^0)$ doublets is then of the form of
Eq. (1), but with the following constraints:
\begin{equation}
\mu_1^2 = m_1^2 + \left( f^2 - {1 \over 4} G_1^2 \right) v_2^2,~~\mu_2^2 =
m_3^2 + \left( f^2 - {1 \over 4} G_2^2 \right) v_2^2,~~\mu_{12}^2 = -fAv_2,
\end{equation}
and
\begin{eqnarray}
\lambda_1 &=& {1 \over 4} (G_1^2 + G_2^2) - {{(4f^2 - G_1^2)^2} \over
{4(G_1^2 + G_2^2)}},  \\ \lambda_2 &=& {1 \over 2} G_2^2 -
{{(4f^2 - G_2^2)^2} \over {4(G_1^2 + G_2^2)}}, \\ \lambda_3 &=& {1 \over 4}
G_2^2 - {{(4f^2 - G_1^2)(4f^2 - G_2^2)} \over {4(G_1^2 + G_2^2)}}, \\
\lambda_4 &=& f^2 - {1 \over 2} G_2^2,~~~\lambda_5 = 0,
\end{eqnarray}
where the second terms on the right-hand sides of the equations for
$\lambda_{1,2,3}$ come from the cubic interactions of $\sqrt 2$Re$\phi_2^0$.
Assuming that $v_2$ is not many orders of magnitude greater than $v_1$ and
$u_1$, then the running of the couplings is not a significant factor and
we have $g_2 = G_2$ and $g_1^{-2} = G_1^{-2} + G_2^{-2}$.  Hence $G_1^2 =
g_1^2 g_2^2 / (g_2^2 - g_1^2)$ and we obtain Eqs. (3) to (6).

Let $x \equiv \sin^2 \theta_W = g_1^2 / (g_1^2 + g_2^2)$, then the new
coupling $f$ can in principle take on any value in the range
\begin{equation}
0 \le f^2 \le {e^2 \over {4x(1-2x)}}.
\end{equation}
As pointed out earlier, the $f=0$ limit corresponds to the MSSM as it must.
The upper limit, on the other hand, corresponds to that of left-right
symmetry, {\it i.e.} $m_1^2 = m_2^2$ in $V_{soft}$, from which Eq. (8) is
obtained, namely
\begin{equation}
\lambda_1 = 0,~~\lambda_2 = {e^2 \over {2x}} \left[ 1 - {{2x^2} \over
{(1-x)(1-2x)}} \right],~~\lambda_3 = {e^2 \over {4x}} \left[ 1 -
{{2x} \over {1-2x}} \right] = - \lambda_4,~~\lambda_5 = 0.
\end{equation}
Now because there are nonnegligible radiative corrections\cite{rad} due to a
large value of $m_t$,
$\lambda_2$ has a significant additional contribution given by $g_2^2
\epsilon / 4 M_W^2 \sin^4\beta$, where
\begin{equation}
\epsilon = {{3 g_2^2 m_t^4} \over {8 \pi^2 M_W^2}} \ln \left( 1 +
{{\tilde m}^2 \over m_t^2} \right).
\end{equation}
In the above, $\tilde m$ is an effective mass for the two scalar
supersymmetric partners of the $t$ quark.  The 2 $\times$ 2 mass-squared
matrix spanning $\sqrt 2$Re$\phi_1^0$ and $\sqrt 2$Re$\phi_2^0$ is then
given by\cite{redef}
\begin{equation}
{\cal M}^2 = \left( \begin{array} {c@{\quad}c} m_A^2 \sin^2\beta & -m_A^2
\sin\beta \cos\beta \\ -m_A^2 \sin\beta \cos\beta & m_A^2 \cos^2\beta +
2 \lambda_2 v_2^2 + \epsilon / \sin^2\beta \end{array} \right),
\end{equation}
where
\begin{equation}
m_A^2 = {{-\mu_{12}^2} \over {\sin\beta \cos\beta}}.
\end{equation}
This implies
\begin{equation}
m_{H_2^0}^2 \leq m_A^2 \sin^2\beta,
\end{equation}
as well as
\begin{equation}
m_{H_2^0}^2 \leq 2 M_W^2 \left[ 1 - {{2x^2} \over {(1-x)(1-2x)}} \right]
\sin^4\beta + \epsilon,
\end{equation}
where $H_2^0$ is the lighter of the two mass eigenstates.  Recall in the
MSSM, the above two bounds are $m_A^2 \cos^2 2\beta + \epsilon/\tan^2\beta$
and $M_Z^2 \cos^2 2\beta + \epsilon$ instead respectively.  Note
that at tree level, $m_{H_2^0} \leq m_A$ is required in both models, but
with radiative corrections, it holds only in this model.  We plot in Figs. 1
and 2 the maximum allowed value of $m_{H_2^0}$ and the minimum allowed value of
$m_{H_1^0}$ as functions of $m_A$ in the MSSM and in this model respectively.
In the limit $m_A = 0$, we have $m_{H_2^0}^2 \leq M_Z^2$ and $m_{H_1^0}^2 \geq
M_Z^2 + \epsilon$ in the MSSM, whereas $m_{H_2^0} = 0$ and $m_{H_1^0}^2
\geq 2 [2 M_W^2 (1 - 2 x^2/(1-x)(1-2x)) \epsilon]^{1 \over 2}$ in this model.
In the limit $m_A$ much greater than the electroweak energy scale, both models
reduce to the standard model with $H_2^0$ as its one physical Higgs boson
such that $m_{H_2^0}^2 \leq M_Z^2 + \epsilon \simeq (115\ {\rm GeV})^2$
 in the MSSM and
$m_{H_2^0}^2 \leq 2 M_W^2 (1 - 2 x^2/(1-x)(1-2x)) + \epsilon \simeq
(120\ {\rm GeV })^2 $
in this model, where we have assumed $m_t$ = 150 GeV and $\tilde m$ = 1 TeV
in estimating the value of $\epsilon$. If we now
consider the decay $Z \rightarrow H_2^0 A$, its experimental nonobservation
at LEP down to the level of $10^{-6}$ in branching fraction restricts the
parameter space of $(m_A, \tan\beta)$ as shown in Fig. 3 for both the MSSM
($f$=0) and this model ($f = f_{max}$).  As for the charged Higgs boson,
the well-known sum rule $m_{H^\pm}^2 = m_A^2 + M_W^2$ in the MSSM becomes
\begin{equation}
m_{H^\pm}^2 = m_A^2 + {1 \over 2} M_W^2 \left( 1 - {{2x} \over {1-2x}} \right)
\end{equation}
in this model.  More details regarding the phenomenological implications of
this model in comparison with the MSSM will be given elsewhere.\cite{inprep}

In conclusion, we have shown in this paper that the requirement of
supersymmetry does not uniquely determine the self-interaction structure
(and thus the mass spectrum) of the two Higgs doublets at the electroweak
energy scale.  The reason is that there may be one or more terms in the
superpotential linking the two Higgs-doublet superfields with a heavier
superfield, as allowed by a larger symmetry at a higher mass scale.  The
reduced Higgs potential at the electroweak energy scale remembers these
couplings and would only be identical to that of the minimal supersymmetric
standard model (MSSM) if these additional couplings were zero.  [The hidden
assumption of the MSSM is in fact the absence of such couplings.]

We show in particular how an especially interesting version\cite{mbhm} of
the supersymmetric $\rm SU(2)_L \times SU(2)_R \times U(1)$ model would
result in two Higgs doublets whose quartic self-couplings are given by
Eqs. (3) to (6), rather than by Eq. (2).  For illustration, we specialize
to the case $f = f_{max} = e/2\sqrt{x(1-2x)}$ and discuss how the Higgs mass
spectrum of this model differs from that of the MSSM.  If future experiments
confirm the existence of two and only two Higgs doublets at the
electroweak energy scale, it will be very important to know whether they are
consistent with supersymmetry and we should bear in mind that the MSSM is not
the only possibility.
\vspace{0.3in}
\begin{center} {ACKNOWLEDGEMENT}
\end{center}

The work of E.M. was supported in part by the U. S. Department of Energy
under Contract No. DE-AT03-87ER40327.  The work of D.N. was supported by
the Natural Sciences and Engineering Research Council of Canada.

\newpage
\bibliographystyle{unsrt}

\begin{thebibliography} {99}
\bibitem{hk} For a review, see for example H. E. Haber and G. Kane,
{\em Phys. Rep.} {\bf 117}, 75 (1985).
\bibitem{bcpetc} See for example V. Barger, K. Cheung, R. J. N. Phillips,
and A. L. Stange, {\em Phys. Rev.} {\bf D46}, 4914 (1992); H. Baer, M.
Bisset, D. Dicus, C. Kao, and X. Tata, {\em ibid.} {\bf D47}, 1062 (1993).
\bibitem{mn} E. Ma and D. Ng, Univ. of Calif., Riverside Report No. UCRHEP-
T103 (1993).
\bibitem{mbhm} E. Ma, {\em Phys. Rev.} {\bf D36}, 274 (1987); K. S. Babu,
X.-G. He, and E. Ma, {\em ibid.} {\bf 36}, 878 (1987).
\bibitem{rad} Y. Okada, M. Yamaguchi, and T. Yanagida, {\em Phys. Lett.}
{\bf B262}, 54 (1991); {\em Prog. Theor. Phys.} {\bf 85}, 1 (1991); H. E.
Haber and R. Hempfling, {\em Phys. Rev. Lett.} {\bf 66}, 1815 (1991); J.
Ellis, G. Ridolfi, and F. Zwirner, {\em Phys. Lett.} {\bf B257}, 83 (1991);
R. Barbieri, M. Frigeni, and F. Caravaglios, {\em ibid.} {\bf 258}, 167
(1991); A. Yamada, {\em ibid.} {\bf 263}, 233 (1991).
\bibitem{redef} We now redefine $(\eta_1^+,\eta_1^0)$ as $(\phi_2^+,\phi_2^0)$.
We also neglect other radiative-correction terms which are proportional to
only $m_t^2$.
\bibitem{inprep} E. Ma and D. Ng, in preparation.

\end{thebibliography}

\newpage

\section*{Figure captions}

\newcounter{Figures}

\begin{list}
{{\bf Fig. \arabic{Figures}.}}
{\usecounter{Figures}}
\item
Maximum value of $m_{H^0_2}$ (solid line) and
minimum value of $m_{H^0_1}$ (dash line) as functions of $m_A$ in
the MSSM.
\item
Maximum value of $m_{H_2^0}$ (solid line) and
minimum value of $m_{H_1^0}$ (dash line) as functions of $m_A$ in
this model.  See Eq. (23) of text.
\item
Contour plots for Br(Z $\rightarrow A H^0_2)=10^{-6}$ as functions of $m_A$
and $\tan \beta$ in the MSSM (dash line)
and in this model (solid line).  The allowed regions are to the right of
the lines.
\end{list}

\end{document}